\documentclass[fleqn,twoside]{article}
\usepackage{espcrc2}

\usepackage{epsfig}

\usepackage{graphicx}
\usepackage[figuresright]{rotating}

\newcommand{\AmS}{{\protect\the\textfont2
  A\kern-.1667em\lower.5ex\hbox{M}\kern-.125emS}}

\title{Pad$\acute{e}$/renormalization-group improvement of inclusive semileptonic B decay rates}

\author{M. R. Ahmady\address{Department of Physics, Mount Allison University, Sackville, NB  E4L 1E6, Canada},
\        F. A. Chishtie\address{F. R. Newman Laboratory of Elementary-Particle Physics,
    Cornell University, Ithaca, NY  14853, USA},
        V. Elias\address{Perimeter Institute for Theoretical Physics,
    35 King Street North, Waterloo, ON  N2J 2W9, Canada},
    \thanks{Permanent Address: Department of Applied Mathematics,
        University of Western Ontario, London, ON  N6A 5B7, Canada}
A. H. Fariborz\address{Department of Mathematics/Science, State
University of New York Institute of Technology, Utica, NY
13504-3050, USA}, D.G.C. McKeon\address[DMP]{Department of
Mathematical Physics, National University of Ireland, Galway,
Ireland},
    \thanks{Permanent Address:  Department of Applied Mathematics,
    University of Western Ontario, London, ON  N6A 5B7, Canada}
    T. N. Sherry\addressmark[DMP]
        and
T. G. Steele\address{Department of Physics and Engineering
Physics, University of Saskatchewan, Saskatoon, SK  S7N 5E2,
Canada}}

\begin{document}

\begin{abstract}
Renormalization Group (RG) and optimized
Pad$\acute{e}$-approximant methods are used to estimate the
three-loop perturbative contributions to the inclusive
semileptonic $b \to u$ and $b \to c$ decay rates. It is noted that
the $\overline{MS}$ scheme works favorably in the $b \to u$ case
whereas the pole mass scheme shows better convergence in the $b
\to c$ case. Upon the inclusion of the estimated three-loop
contribution, we find the full perturbative decay rate to be
$192\pi^3\Gamma(b\to u\bar\nu_\ell\ell^-)/(G_F^2| V_{ub}|^2) =
2065 \pm 290{\rm GeV^5}$ and $192\pi^3\Gamma(b\to
c\ell^-\bar\nu_\ell)/(G_F^2|V_{cb}|^2)= 992  \pm 198 {\rm GeV^5}$,
respectively. The errors are inclusive of theoretical
uncertainties and non-perturbative effects. Ultimately, these
perturbative contributions reduce the theoretical uncertainty in
the extraction of the CKM matrix elements $|V_{ub}|$ and
$|V_{cb}|$ from their respective measured inclusive semileptonic
branching ratio(s). \vspace{1pc}
\end{abstract}

\maketitle The total inclusive rates for semileptonic $b \to c$
and $b \to u$ processes have been calculated to two-loop order in
QCD \cite{CKM99,TVR99}:
\begin{eqnarray}
\Gamma_{bc}/\kappa & = & \left[ m_b \right]^5 F\left(\frac{m_c^2}{m_b^2}\right) \left [ 1 - 1.67x(\mu) \right.\nonumber\\
& - & (8.9 \pm 0.3 + 3.48L(\mu))x^2(\mu)]
\end{eqnarray}
\begin{eqnarray}
\Gamma_{bu}/\kappa & = & \left[ m_b (\mu)\right]^5 \left[ 1 + (4.25360 + 5L(\mu))x(\mu) \right.\nonumber\\
& + & (26.7848 + 36.9902L(\mu) \nonumber\\
& + & \left. 17.2917L^2(\mu))x^2(\mu)\right]
\end{eqnarray}

where
\begin{eqnarray}
x(\mu) \equiv \alpha_s (\mu) / \pi, \; \; \kappa \equiv G_F^2
|V_{ub}|^2 / 192\pi^3,\nonumber\\
F(r)=1-8r-12r^2\log(r)+8r^3-r^4, \nonumber\\
( \hbox{for} \; b \to c) :L(\mu) \equiv \log(\mu^2 / m_b m_c),\nonumber\\
( \hbox{for} \; b \to u) :L(\mu) \equiv \log(\mu^2 / m_b^2 (\mu)).
\end{eqnarray}

In \cite{MRA00} we note that the $b \to u$ rate has less
renormalization scale dependence in the $\overline{MS}$ scheme
than in the pole-mass scheme. However, in the case of $b \to c$,
we find that the total inclusive rate expressed in terms of the
$b$ and $c$ pole masses is better behaved \cite{MRA02}. In both
cases the scale dependence, which is considerable, provides no
optimal choice of renormalization scale $\mu$. Consequently, we
estimate the three-loop contributions to the above rate(s) using
Pad$\acute{e}$ approximants in order to reduce theoretical
uncertainties like scale dependence and truncation error. Both
decay rates have the following general (perturbative
scale-sensitive) form in powers of the strong coupling:
\begin{equation}
S(x)=1+R_1x+R_2x^2+R_3x^3+\ldots \;,
\end{equation}
where $R_1$ and $R_2$ are known as indicated in (1) and (2) . The
(unknown) three-loop contribution term $R_3$ is necessarily of the
form:
\begin{equation}
R_3=c_0+c_1L(\mu)+c_2L^2(\mu)+c_3L^3(\mu).
\end{equation}
Since the decay rate is renormalization group (RG) invariant, the
following relation holds:
\begin{equation}
\mu^2\frac{\mathrm{d}\Gamma}{\mathrm{d}\mu^2}=0.
\end{equation}
This allows us to evaluate $c_1$, $c_2$ and $c_3$ exactly for both
the decay rates. For the case of $b \to c$ we obtain
\begin{equation}
c_1=-42.4\pm 1.3~,~c_2= -7.25~,~c_3=0
\end{equation}
and for the $b \to u$ case we obtain:
\begin{equation}
c_1=249.6,~c_2= 178.8~,~c_3=50.9
\end{equation}
The estimate for the RG-inaccessible coefficient $c_0$ is obtained
from the following estimate developed via asymptotic
Pad$\acute{e}$ approximant methods \cite{MRA02}:
\begin{equation}
R_3^{Pade}=\frac{(2+k)R_2^3}{(1+k)R_1^3+R_1R_2}
\end{equation}
The quantity $k$ parametrizes a family of Pad$\acute{e}$
approximants as outlined in \cite{MRA02}. Since both the assumed
form of the three-loop contribution and Pad$\acute{e}$-estimated
version are dependent on $\mu$, we proceed by evaluating the
following moments:
\begin{equation}
N_j=(j+2)\int\limits_0^1w^{j+1}R_3(w)\mathrm{d}w
\end{equation}
where $\log w =-L(\mu)$. Hence, to estimate the value for $c_0$,
we match the scale dependence of the known form (5) to the
Pad$\acute{e}$ estimate (9), using the first four moments
($N_{-1}$, $N_0$, $N_1$, $N_2$) in the perturbative (UV) region.
This leads to four linear equations for the four three-loop
coefficients \{$c_0$, $c_1$, $c_2$, $c_3$\}. For the $b \to u$
case the method works quite well with $k = 0$ \cite{MRA00}.
However, in the case of $b \to c$ decay rate (pole mass scheme)
the estimate obtained from $k = 0$ is ill-suited to the series in
question \cite{MRA02}. A value of $k$ for the $b \to c$ case can
be obtained by finding an optimal $k$ which minimizes the sum of
the squares of the relative errors in predicting $c_1$ and $c_2$,
as given in (7).

Applying the above procedure, we get the following estimated
values of $c_{0-3}$ for the $b \to c$ case ($k=-0.94$ for the
central value of $b_0 = -8.9$):
\begin{eqnarray}
c_3=2.0\times10^{-4}, c_2=-7.68, \ c_1=-39.7, \nonumber\\
c_0=-51.2
\end{eqnarray}
The above estimates of the RG-accessible coefficients $c_1$ and
$c_2$ are within 7\% error of their correct values (7).

As a consistency check, we substitute the exact RG values into the
moment equations to evaluate $c_0$. We then find that
\begin{eqnarray}
c_0=N_{-1}-c_1-2c_2-6c_3=-49.4~,\nonumber\\
c_0=N_0-\frac{1}{2}c_1-\frac{1}{2}c_2-\frac{3}{4} c_3=-50.1~,\nonumber\\
c_0=N_1-\frac{1}{3}c_1-\frac{2}{9}c_2-\frac{2}{9} c_3=-50.4~,\nonumber\\
c_0=N_2-\frac{1}{4}c_1-\frac{1}{8}c_2-\frac{3}{32} c_3=-50.6~.
\end{eqnarray}which are all within 3\% of the Pad$\acute{e}$-estimated value in Eq.(11).

Similarly, for the $b \to u$ case we have
\begin{eqnarray}
c_3=47.61, c_2=190.5, \ c_1=251.4,\nonumber\\c_0=206
\end{eqnarray}
The RG-accessible coefficients $c_1$, $c_2$ and $c_3$ are all
within 7\% of their true values (8). A similar consistency check
(as above) gets us values of $c_0$ within 3\% of the
Pad$\acute{e}$-estimated value in Eq.(13).

Upon inclusion of the 3-loop contribution we see decreased
renormalization scale dependence and the emergence of PMS
(Principle of Minimal Sensitivity \cite{PMS81}) extrema for the
decay rates \cite{MRA02,MRA00}. For the $b \to c$ case, we find
that the PMS value occurs at $\mu$ = 1 GeV and yields
$\Gamma_{bc}/\kappa$ = 1047 GeV$^5$. We find that this is
remarkably close to the FAC (Fastest Apparent Convergence
\cite{GG80}) value which occurs at $\mu$ = 1.18 GeV and yields
$\Gamma_{bc}/\kappa$ = 1051 GeV$^5$. Similarly, for the $b \to u$
case, we find that the PMS and FAC values occur at $\mu$ = 1.775
GeV and 1.835 GeV, respectively. The corresponding reduced rates,
are 2069 GeV$^5$ and 2071 GeV$^5$, respectively, which are
virtually identical.

We take the PMS value of both the decay rates as our central
value. Further, we assume that the error in estimating $c_0$ is
the same as our largest error in estimating an RG-accessible
coefficient for both cases. We then obtain the following value for
the total decay rate for $b \to c$ case:
\begin{equation}
\Gamma_{bc}/\kappa  = 992  \pm 198  {\rm  GeV^5}.
\end{equation}
This error estimate includes the uncertainties in the two-loop
contribution ($b_0$), the $b$-quark pole mass ($m_b=4.9$ $\pm$
$0.1$ \cite{AH00}), the strong coupling constant, the three loop
estimate, and the uncertainty in estimating non-perturbative
effects (for details see \cite{MRA02}). One can extract $|V_{cb}|$
from the above expression with a theoretical error of 10\%.

Using the same set of uncertainties (note that $m_b(m_b)=4.17 \pm
0.05$ \cite{KGC81},\cite{MRA00}) we obtain the following value for
the $b \to u$ rate:
\begin{equation}
\Gamma_{bu}/\kappa  = 2065  \pm 290  {\rm  GeV^5}
\end{equation}
The above expression implies that from the total decay rate one
can extract $|V_{ub}|$ with a theoretical error of 7\%. However,
in this case, the experimental outlook on obtaining a comparably
precise measurement of the total inclusive rate (due to the
presence of an enormous charm background) does not look promising
at present.

We are grateful for support from the Natural Sciences and
Engineering Research Council of Canada.

\end{document}